\documentclass{article}
\usepackage{amsmath,amsthm,amssymb,amsfonts}
\usepackage{mathabx}
\usepackage{fullpage}
\usepackage{graphicx}
\usepackage{dsfont}
\usepackage{bm}
\usepackage{color}
\usepackage{url}

\newcommand{\e}{\mathrm{e}}

\newcommand{\Var}{\mathop{\mathrm{Var}}}
\newcommand{\eps}{\varepsilon}

\newcommand{\dA}{\mathrm{d}A}
\newcommand{\dx}{\mathrm{d}x}
\newcommand{\dy}{\mathrm{d}y}
\newcommand{\dz}{\mathrm{d}z}
\newcommand{\ddx}{\frac{\mathrm{d}}{\dx}}
\newcommand{\ddz}{\frac{\mathrm{d}}{\dz}}
\newcommand{\dtheta}{\mathrm{d}\theta}

\newcommand{\dt}{\mathrm{d}t}

\renewcommand{\vec}[1]{{\bm{#1}}}
\renewcommand{\i}{\mathrm{i}}

\begin{document}

\title{\emph{Astronomia Ova}:\\ Kepler's Laws, Ptolemy's Equant, and Legendre's Polynomials}
\author{Cristopher Moore\footnote{Santa Fe Institute, Santa Fe, New Mexico. \texttt{moore@santafe.edu}}
\and Malcolm Smith\footnote{The MASTERS program, Santa Fe, New Mexico. \texttt{msmith.malcolmsmith@gmail.com}}
\and Stephan Mertens$^{*,}$\footnote{Otto-von-Guericke University, Magdeburg, Germany.}}
\maketitle

\begin{abstract}
Motivated by Kepler's long battle with the orbit of Mars, we present a family of orbits that satisfy his second law, conserving angular momentum around a central mass, and also fulfill Ptolemy's belief in an equant---a point around which the angular speed is constant. For specific values of the angular momentum, these orbits can be written in terms of Legendre polynomials and  functions of the second kind. They are analytic except at perihelion where they are twice differentiable. They  resemble the ``ovoid'' (egg-shaped) or ``metopoid'' (face-shaped) orbits that Kepler considered on his way to the ellipse.
\end{abstract}

\section{Introduction}

Kepler's discovery of elliptical orbits using Tycho Brahe's observations is one of the great triumphs of empirical science. It is even more extraordinary when we consider that the theory Kepler refuted was itself far from naive. It was a clever theory of nonuniform circular motion, based on careful observation, that fits much astronomical data very well. 

It was known to the Babylonians that the angular speed of the Moon in the sky increases and decreases over the course of a month~\cite[II.B]{neugebauer1975}. Similarly, the Earth's seasons between equinoxes and solstices are not of equal length, showing that the angular speed of the Earth around the Sun, or apparently of the Sun around the Earth, varies over the course of the year. Since the idea of uniform circular orbits was hard to give up at the time, the Greek astronomers Apollonius and Hipparchus addressed these ``anomalies'' in angular speed by proposing that orbits are circular but eccentric, with centers offset from the Earth.\footnote{As Apollonius proved, this is equivalent to an epicycle mounted on a circular ``deferent'' centered on the Earth, as long as the epicycle's rotation is equal and opposite to that of the deferent.} 

\begin{figure}
  \centering
  
\includegraphics[width=2.35in]{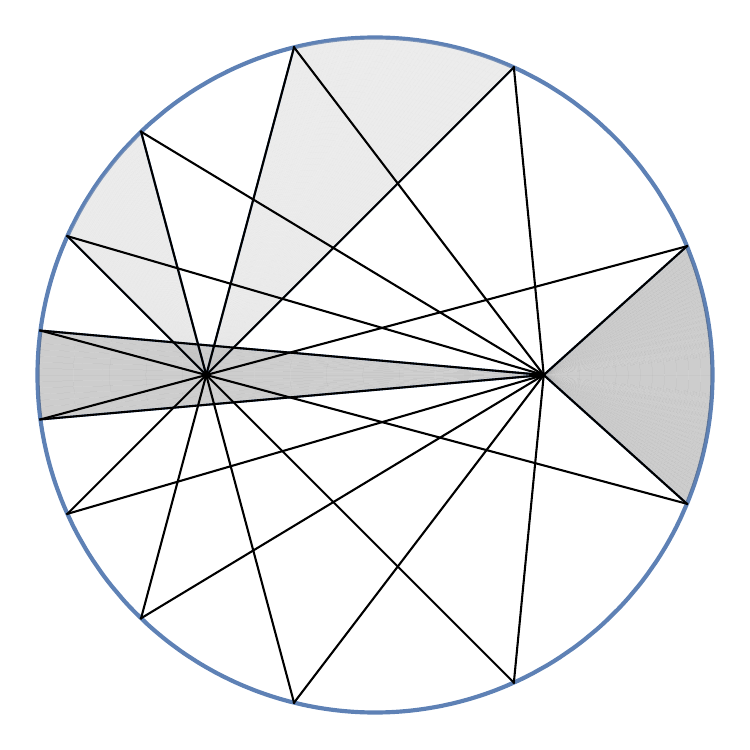}
  \caption{Ptolemy's equant theory. The planet travels in a circular orbit off-center from the Earth, with a  speed that varies so that its angular speed from the point of view of the equant (the point on the left) is exactly constant. As a result, its angular speed from the point of view of the earth (the point on the right) is greater at perihelion than at aphelion, fitting ancient observations. The equant and the Earth are equidistant from the center, an assumption (the ``bisection of the eccentricity'') that Kepler relaxed in his vicarious hypothesis. Each wedge corresponds to an amount of time equal to $1/12$ of the orbital period.}
  \label{fig:ptolemy}
\end{figure}

This eccentric theory was a significant advance, and appears to have been used in the Antikythera device to predict the position of the Moon~\cite{pin-and-slot,price1974gears}. But its quantitative accuracy is not very good, especially for retrograde motion~\cite{evans84}.
In order to ``save the phenomena'' and better fit observation, Ptolemy further proposed that planets do not move on these circles with constant speed, but rather that their angular speed is uniform when viewed from a third point across the center from the Earth as shown on the left in Figure~\ref{fig:ptolemy}. This point was later called the \emph{punctum aequans} or \emph{equant}. If an observer is located at the equant, the planet appears to move at constant speed through the sky.

Ptolemy's theory fits observations quite well, and stood for 1400 years. Indeed, the motions predicted by Ptolemy and Kepler are almost identical as long as the eccentricity $\eps$ of the ellipse is small~\cite{aaboe2001kepler,Brenke-36}. Specifically, the empty focus of the ellipse, across the center from the Sun or Earth as Ptolemy suggested, is an approximate equant: the angular speed of a planet around it is constant up to $O(\eps^2)$ error. This makes Ptolemy's theory highly accurate for planets with nearly-circular orbits like Venus and the Earth, or the Sun in the geocentric view. 

Like Copernicus, Kepler regarded the equant as philosophically suspect. He wanted an astronomy driven by physical causes, even if he personified these causes as ``minds'' and ``virtues''~\cite{stephenson1987}. It is unclear how a heavenly body could ``know'' about a vacant point in space and direct its movement accordingly. But he might have been unable to reject the equant on scientific grounds if not for Mars~\cite{Bracco2009}, whose eccentricity $\eps=0.09$ is the largest of any visible planet except Mercury, the orbit of which is largely obscured by the Sun. Kepler found that Ptolemy's theory predicts Mars' position in the sky within $8'$, about 13\% of a degree or a quarter of the width of the Moon. This is within the accuracy available in antiquity, but Tycho's observations were accurate to within $1'$ or $2'$. Thus Ptolemy's theory had to be abandoned or corrected.

Over five long years Kepler tested many  models for Mars. These included a ``vicarious hypothesis'' where the orbit is circular but the equant can be anywhere on the line of symmetry---generalizing Ptolemy's ``bisection of eccentricity'' where the equant and the Sun are equidistant from the center. But adjusting the equant to fit Mars' angle caused unacceptable errors in its distance from the Sun, and vice versa. Finally Kepler was forced to abandon circular orbits and the equant entirely, which led him to the ellipse.

Well after Kepler published his \emph{Astronomia Nova} in 1609, some astronomers remained fond of the equant. In the mid-1600s, Seth Ward, Isma\"{e}l Boulliau, Thomas Streete, and Emile-Fran\c{c}ois de Pagan used variants of a ``simple elliptic theory'' that had previously been proposed to Kepler by 
Albert Curtz~\cite{wilson70,gaythorpe25}. These astronomers accepted Kepler's first law that orbits are ellipses, but disagreed with or misunderstood his second law of equal area in equal time~\cite{russell64}. In particular, they were frustrated that, to find the position of a planet on the ellipse as a function of time, it was necessary to solve a transcendental equation using iterative methods---a situation they found ungeometric. Instead, they believed the angular speed around the empty focus is constant, making it an exact equant. This gives a simple geometric theory with a pleasing uniform motion, and makes it much easier to locate a planet at a given time. 
But it comes at a cost: namely, the second law only holds approximately. While some like Nicholas Mercator defended the second law on empirical grounds~\cite{wilson70,applebaum}, it was not until the 1680s that Newton showed that it is implied by a central force, and gave Kepler's laws a mechanistic explanation from the inverse-square law.

In this paper we carry out a mathematical thought experiment. For Kepler, the second law is exact but the equant is approximate. In the simple elliptic theory, the second law is approximate but the equant is exact. Can we have both, at the cost of the ellipse? Is there an orbit that  satisfies the second law around one point, and has an equant at another point? 

In what follows, we first review the fact that the empty focus is an approximate equant. Inspired by Kepler's vicarious hypothesis, we then ask whether some other point is an even better equant, with a more uniform angular speed around it. We find that the variance of the angular speed is minimized at a point slightly closer to the center, so that bisecting the eccentricity is not quite optimal in this sense. We also briefly discuss a non-elliptical model with an equant proposed by Cassini~\cite{cassini1693origine,wilson1989predictive}.

We then turn to our main goal: finding an orbit that vindicates Ptolemy at one point and Kepler, or at least his second law, at the other. The resulting differential equations give rise to Legendre polynomials and related functions, and we give analytic solutions for certain values of the angular momentum. These orbits are egg-shaped, with more curvature near the Sun than the equant, like the ``puffy-cheeked'' or ``face-shaped'' ovals Kepler investigated on his way to the ellipse~\cite{whiteside-74,Carman2025-metopoid}. Thus they provide a precise, and perhaps intriguing, mathematization of a side branch on the great tree of physics.

\section{Kepler orbits and the Approximate Equant}
\label{sec:kepler}

\subsection{A little geometry}

\begin{figure}
  \centering
  \includegraphics[width=2.5in]{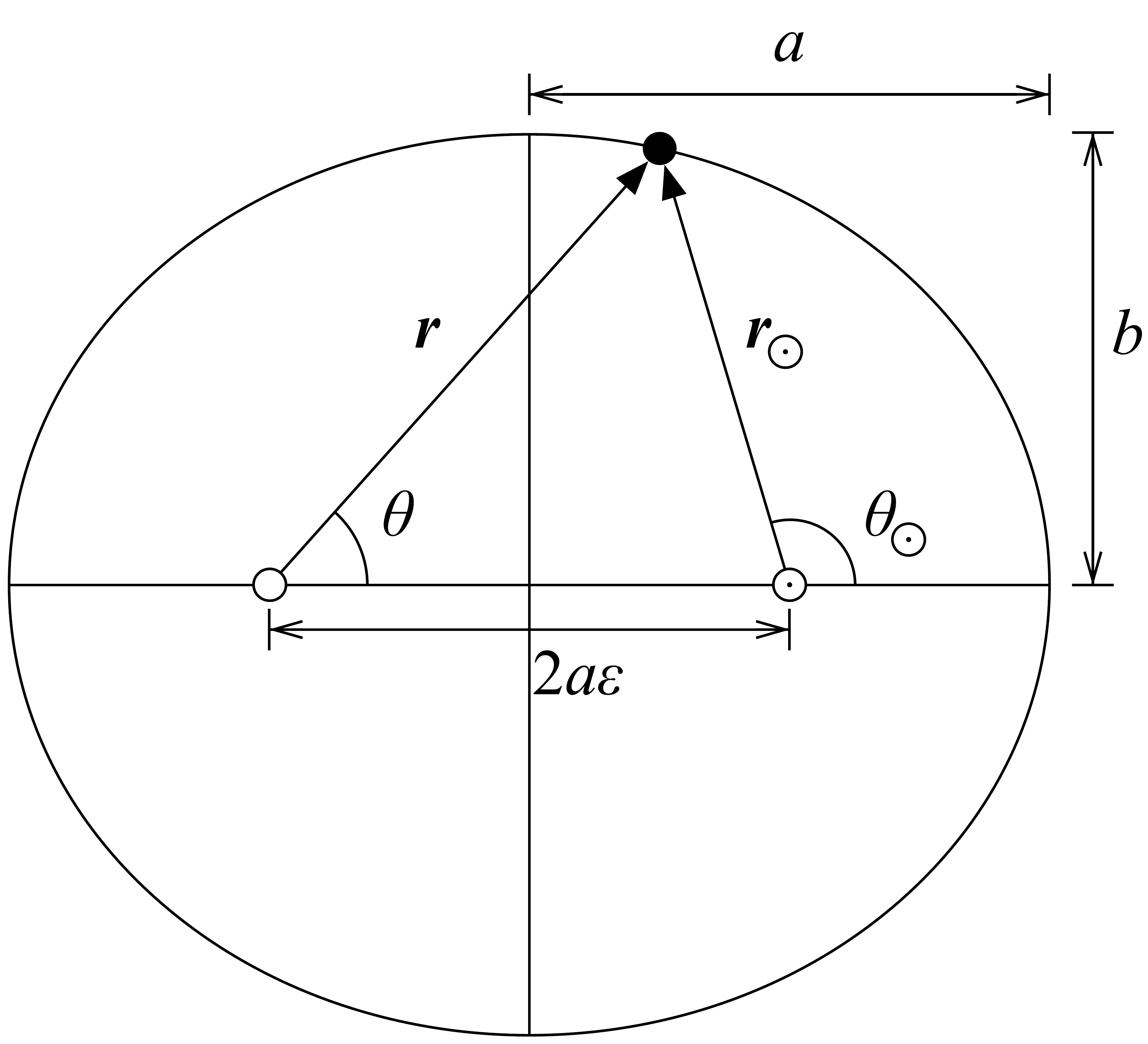}
  \caption{Distances and angles as viewed from the empty focus $\ovoid$, which we treat as the origin, and the sun $\odot$. This figure shows an ellipse with eccentricity $\eps = 1/2$.}
  \label{fig:ellipse-foci}
\end{figure}

The \emph{eccentricity} 
$\eps$ of an ellipse is defined as the ratio between the distance between its foci and its length along the major axis. Consider a planet whose orbit is an ellipse with semimajor axis $a$, semiminor axis $b$, and eccentricity $\eps$, as shown in Figure~\ref{fig:ellipse-foci}. We place the empty focus at the origin, and the other focus---which we call the Sun---at $(2a\eps,0)$. We define $\theta$ and $\theta_\odot$ as the angles above the $x$-axis as seen from the two foci.

Perihelion and aphelion, the moments when the planet is closest to and farthest from the Sun, occur at $\theta=\theta_\odot=0$ and $\theta=\theta_\odot=\pi$, where  
\[
r(0) = a (1+\eps) 
\quad \text{and} \quad
r(\pi) = a (1-\eps) \, . 
\]
More generally, its distance $r$ from the empty focus, and similarly its distance $r_\odot$ from the sun, is given by the equation for an ellipse in polar coordinates,
\begin{equation}
\label{eq:r0}
r(\theta) = \frac{a(1-\eps^2)}{1 - \eps \cos \theta} 
\quad \text{and} \quad 
r_\odot(\theta_\odot) = \frac{a(1-\eps^2)}{1 + \eps \cos \theta_\odot} 
\, .
\end{equation}
where $\theta$ and $\theta_\odot$ as shown in Figure~\ref{fig:ellipse-foci}. If we use the law of cosines to write
\[
r_\odot^2
= 
(2a\eps)^2 + r^2 
- 4a\eps r \cos \theta \, ,
\]
we can confirm that the sum of the distances from the foci is a constant, 
\[
r + r_\odot = 2a \, .
\]
Since $r^2 = r^2_\odot = (a\eps)^2 + b^2 = a^2$ at the top and bottom of the orbit, the semiminor axis is 
\[
b = a \sqrt{1-\eps^2} \, . 
\]



We can find $\dtheta_\odot/\dt$ using Kepler's second law. In an interval $\dt$, the orbit sweeps an infinitesimal triangle of base $r_\odot$, height $r_\odot \,\dtheta_\odot$, and area $r_\odot^2 \,\dtheta_\odot / 2$. Thus the rate at which the orbit sweeps out area is 
\[
\frac{\dA}{\dt} = \frac{r_\odot^2}{2} \frac{\dtheta_\odot}{\dt} \, .
\]
Since Kepler's second law asserts that $\dA/\dt$ is constant, $\dtheta_\odot/\dt$ is proportional to $1/r^2_\odot$. 
Putting this in modern terms, $\dA/\dt$ is proportional to the angular momentum. If the planet has mass $m$ we have (where $J$ is the component of the angular momentum pointing up out of the plane)
\begin{equation}
\label{eq:J-dAdt}
\frac{J}{m} 
= |\vec{r}_\odot \times \vec{v}| 
= r_\odot^2 \frac{\dtheta_\odot}{\dt} 
= 2 \,\frac{\dA}{\dt} \, ,
\end{equation}
so the conservation of angular momentum implies that $\dA/\dt$ is constant. Over the course of its period $T$, the orbit sweeps out the entire area $A=\pi a b$ of the ellipse, so 
\begin{equation}
\label{eq:J-ellipse}
\frac{J}{m} 
= \frac{2\pi a b}{T} 
= \frac{2\pi a^2 \sqrt{1-\eps^2}}{T} \, .
\end{equation}

For a given planet and a given orbit, let us rescale  mass, length, and time by setting $m=1$, $a=1$, and $T=2\pi$. This implies that the mean angular speed around the empty focus, or any other point inside the orbit, is  $\overline{\omega}=1$. Then~\eqref{eq:J-dAdt} and~\eqref{eq:J-ellipse} become
\begin{equation}
\label{eq:after-scaling}
J = \sqrt{1-\eps^2} 
\quad \text{and} \quad
\frac{\dtheta_\odot}{\dt} 
= \frac{J}{r_\odot^2} 
= \frac{\sqrt{1-\eps^2}}{r_\odot^2} 
\, .
\end{equation}
We caution the reader that unlike, say, the classic choice of setting $c=h=G=1$ in relativity and cosmology, this does not correspond to a consistent choice of units. These length and time scales depend on the planet's initial position and velocity and on the mass of the Sun. Nevertheless, this rescaling will let us focus on the  geometry of Kepler orbits of varying eccentricities, and look for corresponding geometries in our fictional universe.



\subsection{Angular motion around the empty focus}

In our chosen scaling, the angular speed $\dtheta_\odot/\dt$ around the Sun is given by~\eqref{eq:after-scaling}. Now let's consider $\dtheta/\dt$, the angular speed around the empty focus at the origin. 

\begin{figure}
  \centering
  \includegraphics[width=2.5in]{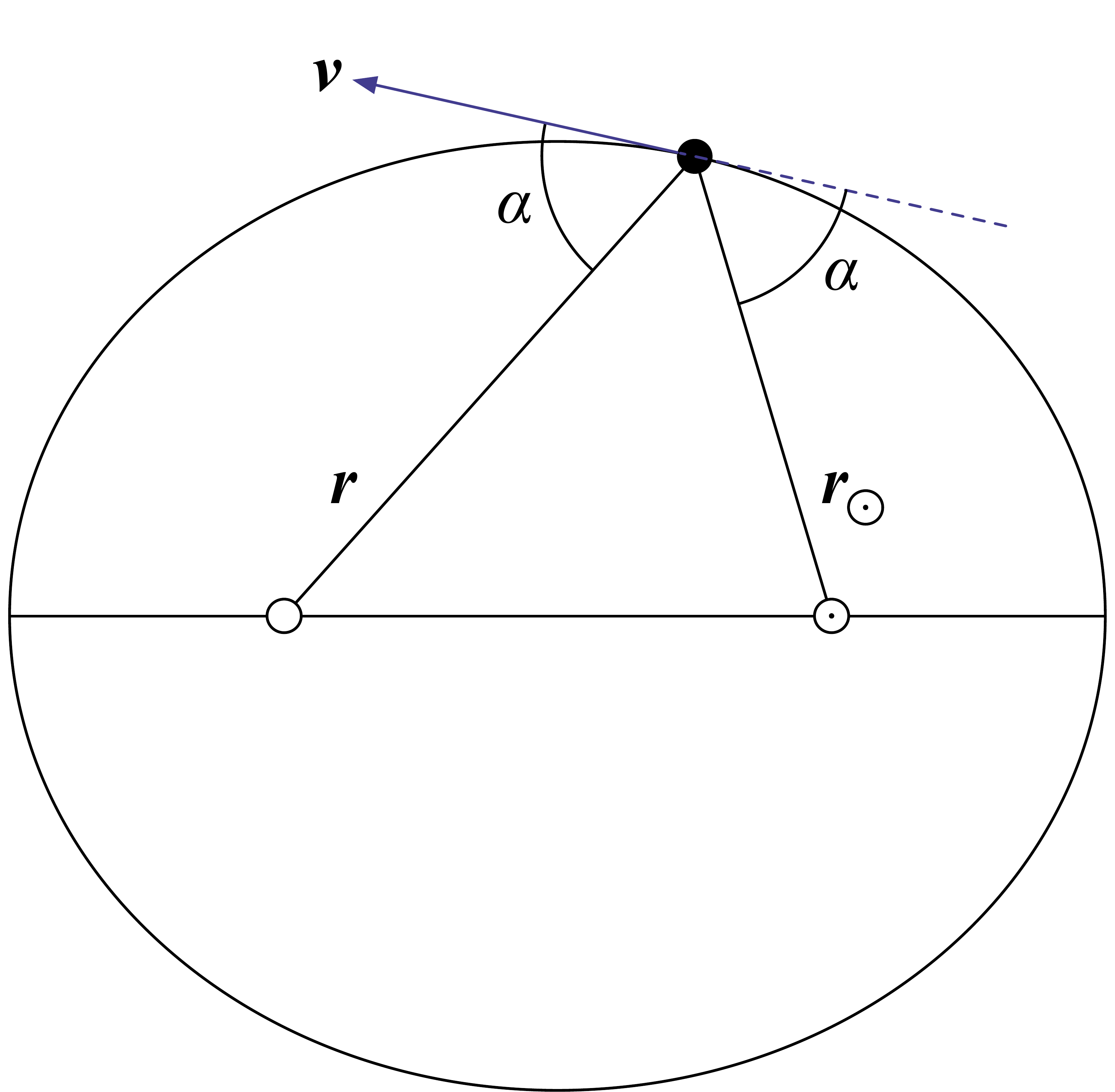}
  \caption{The reflection principle. Both $\vec{r}$ and $\vec{r}_\odot$ form the same angle $\alpha$ with a tangent to a given point on the ellipse, so the transverse component of the velocity is $v \sin \alpha$ when viewed from either focus.}
  \label{fig:reflection}
\end{figure}

The reflection principle states that an ellipse will reflect light rays sent from one focus toward the other. As a result, a tangent to the ellipse  forms the same angle with both $\vec{r}_\odot$ and $\vec{r}$, and thus the transverse component of $\vec{v}$ is the same from both points of view (see Figure~\ref{fig:reflection}). This gives
\begin{equation}
\label{eq:reflection}
r \frac{\dtheta}{\dt} = r_\odot \frac{\dtheta_\odot}{\dt} \, ,
\end{equation}
so~\eqref{eq:after-scaling} gives
\begin{equation}
\label{eq:rr}
\frac{\dtheta}{\dt} 
= \frac{r_\odot}{r} \frac{\dtheta_\odot}{\dt} 
= \frac{\sqrt{1-\eps^2}}{r r_\odot} \, . 
\end{equation}
If $r r_\odot$ were constant then $\dtheta/\dt$ would be constant, and the empty focus would qualify as Ptolemy's equant. This is not true, but it is \emph{almost} true. Since $r$ and $r_\odot$ have a constant sum, if they are not too different from each other---that is, if the eccentricity is not too large---they have an almost constant product as well. We can see this in Figure~\ref{fig:angles} below, where wedges corresponding to equal time intervals have equal areas at the sun, and roughly equal angular width at the empty focus. 

\begin{figure}
  \centering
  
\includegraphics[height=2.5in]{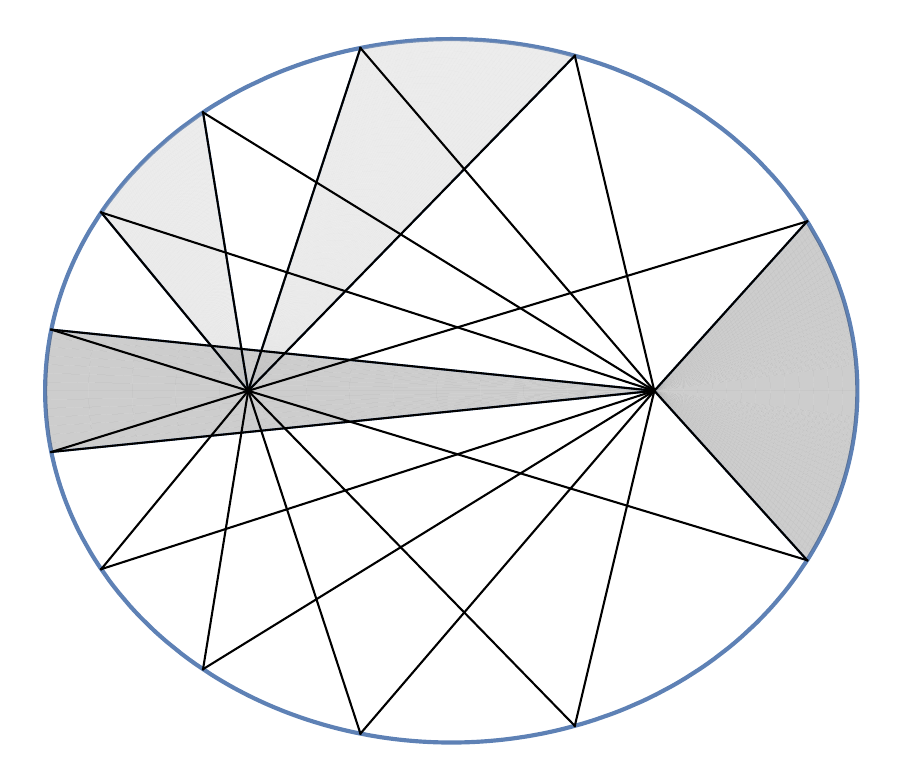}
  \caption{If we draw wedges corresponding to equal time intervals according to Kepler's laws, those drawn from the sun have equal area, and those drawn from the empty focus have approximately equal angular width. Here $\eps=1/2$. As in Figure~\ref{fig:ptolemy}, each wedge corresponds to an amount of time equal to $1/12$ of the orbital period. In Ptolemy's theory, the angular speed around the equant is exactly constant; to the eye, the same seems almost true here.}
  \label{fig:angles}
\end{figure}

The angular speed is maximized when the product $rr_\odot$ is minimized. This occurs at perihelion and aphelion, where (recalling $a=1$) 
\[
rr_\odot 
= (1+\eps)(1-\eps) 
= 1-\eps^2 
\quad \text{and} \quad
\frac{\dtheta}{\dt} = \frac{1}{\sqrt{1-\eps^2}} \, . 
\] 
Indeed, the success of the equant hypothesis is largely due to the fact that $\dtheta/\dt$ is the same at these two extremes of the orbit, where many ancient astronomical measurements were made. 

Similarly, $\dtheta/\dt$ is minimized when $rr_\odot$ is maximized. This occurs when the planet crosses the minor axis, where
\[
rr_\odot = r^2 = 1 
\quad \text{and} \quad 
\frac{\dtheta}{\dt} = \sqrt{1-\eps^2} 
\, . 
\]
Thus $\dtheta/\dt$ ranges between two values, with a ratio of $1-\eps^2$ between them. So the angular speed around the empty focus is uniform up to an error of $O(\eps^2)$, making it an approximate equant.

A smoother way to quantify the nonuniformity of $\dtheta/\dt$ is its variance. Recall that its mean $\overline{\omega}$ is $1$ since a full revolution takes time $T=2\pi$. Its variance is then
\[
\Var \frac{\dtheta}{\dt} = -1 + \frac{1}{2\pi} \int_0^{2\pi} \dt \left( \frac{\dtheta}{\dt} \right)^{\!2} \, . 
\]
Using~\eqref{eq:reflection} and~\eqref{eq:after-scaling} and changing variables gives
\begin{align*}
\Var \frac{\dtheta}{\dt} 
&= -1 + \frac{1}{2\pi} \int_0^{2\pi} \dt \left( \frac{\dtheta_\odot}{\dt} \right)^{\!2} \left( \frac{r_\odot}{r} \right)^{\!2} \\
&= -1 + \frac{1}{2\pi} \int_0^{2\pi} \dtheta_\odot \frac{\dtheta_\odot}{\dt} \left( \frac{r_\odot}{r} \right)^{\!2} \\
&= -1 + \frac{\sqrt{1-\eps^2}}{2\pi} \int_0^{2\pi} \frac{\dtheta_\odot}{r^2} \\
&= -1 + \frac{\sqrt{1-\eps^2}}{2\pi} \int_0^{2\pi} \frac{\dtheta_\odot}{(2-r_\odot)^2} \, , 
\end{align*}
which after some calculation gives
\begin{equation}
\label{eq:var-f2}
\Var \frac{\dtheta}{\dt} 
= -1 + \frac{2-\eps^2}{2 \sqrt{1-\eps^2}}
= \frac{\eps^4}{8} + \frac{\eps^6}{8} +  O(\eps^8) 
\, .
\end{equation}
Since the variance of $\dtheta/\dt$ is $O(\eps^4)$, its standard deviation is $O(\eps^2)$, the same order we obtained above for the gap between its minimum and maximum values.


This leaves us with the question: is the empty focus the most uniform point? That is, the viewpoint inside the orbit that minimizes the variance in angular speed?

\subsection{Angular motion from other viewpoints}

Ptolemy assumed that the center of the orbit was midway from the sun to the equant---what he called the ``bisection of eccentricity.'' In his vicarious hypothesis, Kepler relaxed this assumption by considering equants elsewhere on the line between the sun and the center. Inspired by Kepler's attempts to reconcile Ptolemy with Brahe's observations,  
let's consider the angular motion from a viewpoint $(\ell,0)$ on the major axis, recalling that the empty focus and the sun are at $\ell=0$ and $\ell=2\eps$ respectively. 

Let $\vec{r}_\ell$ be the displacement from this viewpoint to the planet, and let $\theta_\ell$ be the corresponding angle. Then
\[
\frac{\dtheta_\ell}{\dt} 
= \frac{|\vec{r}_\ell \times \vec{v}|}{r_\ell^2} 
\quad \text{where} \quad
\vec{r}_\ell = (r \cos \theta - \ell, r \sin \theta) 
\quad \text{and} \quad
r_\ell^2 = |\vec{r}_\ell|^2 = r^2 + \ell^2 - 2 r \ell \cos \theta \, ,
\]
where $\theta$ is again the angle from the empty focus at $\ell=0$ and  $r_\ell = |\vec{r}_\ell|$. 

As before, since $T=2\pi$ the mean of $\dtheta_\ell/\dt$ is  $1$ as long as $(\ell,0)$ is in the interior of the ellipse, i.e., if $\eps - 1 < \ell < \eps + 1$. Its variance is then 
\[
\Var \frac{\dtheta_\ell}{\dt} = -1 + \frac{1}{2\pi} \int_0^{2\pi} \dt \left( \frac{\dtheta_\ell}{\dt} \right)^{\!2} \, . 
\]
Since $\theta$ is not an elementary function of $t$, we found it easier to change variables, and write
\[
\Var \frac{\dtheta_\ell}{\dt}
= -1 + \frac{1}{2\pi} \int_0^{2\pi} \dtheta \,\frac{\dtheta}{\dt} \left( \frac{\dtheta_\ell}{\dtheta} \right)^{\!2}
\]

The resulting integral is admittedly complicated. After a lengthy computer-assisted calculation we obtain the following expression:
\begin{equation}
\label{eq:var-exact}
W(\ell) := 
\Var \frac{\dtheta_\ell}{\dt} 
= -1 + 
\frac{4 
(1 - \eps^2) 
(2 - \eps^2) 
+ 4 (2 - \eps^2) \eps \ell
- (4 - \eps^2 + \eps^4) \ell^2 
+ \eps^3 \ell^3}
 {2 \sqrt{1-\eps^2} 
 \,(2-\eps \ell)^2 
 \,(1 - (\eps-\ell)^2)}
   \, .
\end{equation}
As a sanity check, $W(\ell)$ recovers the variance~\eqref{eq:var-f2} around the empty focus if we set $\ell=0$. It also has poles at $\ell = \eps \pm 1$ where the viewpoint touches the orbit at aphelion or perihelion and $\dtheta_\ell/\dt$ diverges. 

\begin{figure}
  \centering
  \includegraphics[width=6.5in]{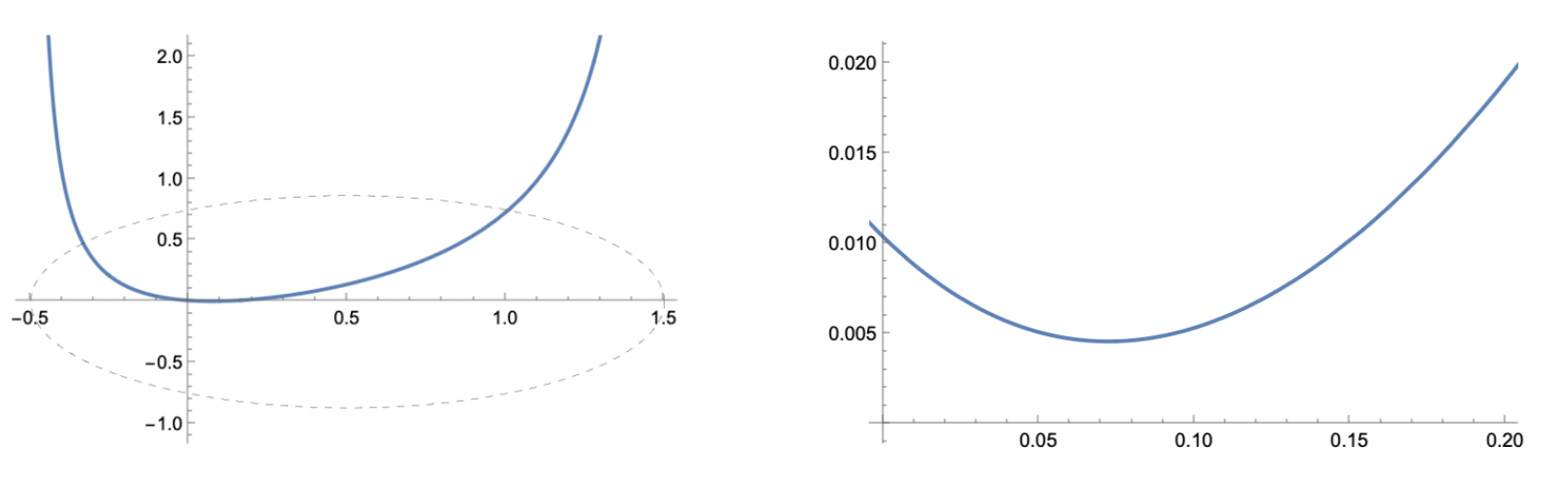}
  \caption{The variance of motion is minimized about a point slightly to the right of the empty focus.}
  \label{fig:ellipse-var}
\end{figure}

We graph $W(\ell)$ 
for an orbit of eccentricity $\eps=1/2$ in Figure~\ref{fig:ellipse-var}. We can see that it is minimized close to the empty focus at $\ell=0$. But zooming in, we see that its true minimum is slightly closer to the center of the ellipse, at $\ell=0.07$. Is the empty focus at least an approximate minimum? 

If we try to minimize $W(\ell)$ exactly, we obtain a fifth-order polynomial in $\ell$. However, we can approximate the relevant root $\ell_{\min}$ of this polynomial by taking Taylor series around the empty focus. Expanding to second order in $\ell$, we obtain
\begin{equation}
\label{eq:vicarious}
\ell_{\min} = 
-\frac{W'(0)}{W''(0)} = \eps^3 + O(\eps^5) 
\quad \text{and} \quad
W(\ell_{\min}) = \frac{\eps^4}{8} - \frac{3 \eps^6}{8} + O(\eps^8) \, . 
\end{equation}
Thus the variance is minimized at a point about $\eps^3$ to the right of the empty focus, slightly closer to the sun. Comparing to~\eqref{eq:var-f2}, we see that the minimum variance is about $\eps^6/2$ smaller than that at the empty focus---a very slight improvement.

\subsection{The Cassinoid}

In~\cite[p.36]{cassini1693origine}, Giovanni Domenico Cassini proposed an alternative to the ellipse. Perhaps inspired by~\eqref{eq:rr} or something like it, he wrote
\begin{quote}
    \ldots a line has been devised to serve as an orbit for the planets. This line is a type of ellipse in which the rectangles [products] formed by the lines [distances] drawn from the planet to both foci are always equal; whereas in ordinary ellipses, it is the sums of the two distances from the foci that are always equal to each other.
\end{quote}
The resulting quartic curve, now called a \emph{cassinoid}, was also discussed by his son Jacques~\cite[pp.149--151]{jacques-cassini}. 

\begin{figure}
  \centering
  \includegraphics[width=2.5in]{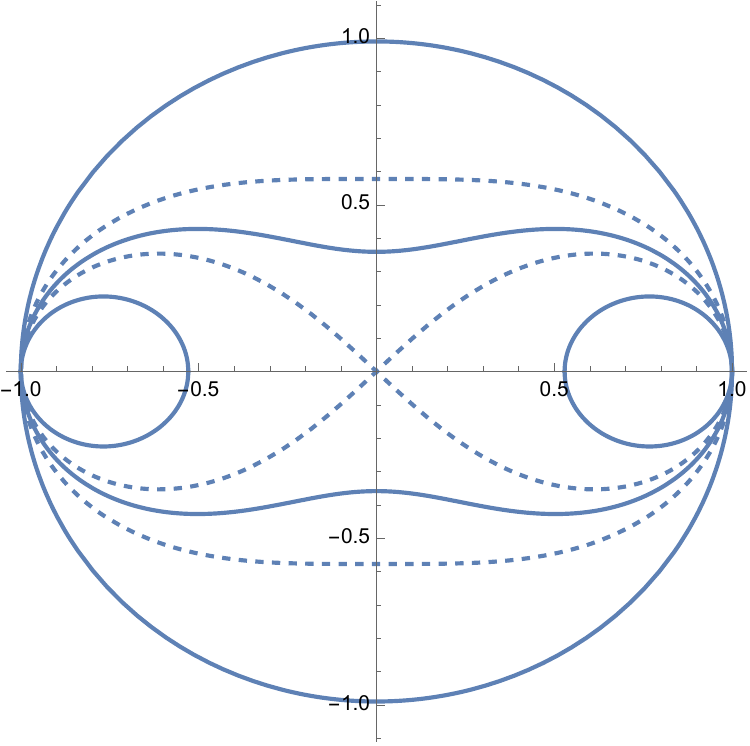}
  \caption{Cassinoid ovals where the product, rather than the sum, of the distances to the foci is a constant. Here we hold the distance between perihelion and aphelion constant while increasing the eccentricity $\eps$. We show $\eps=0.1$ where the cassinoid is close to the ellipse, $\eps=1/\sqrt{3}$ (dashed) where it ceases to be convex, $\eps=0.66$ where it is peanut-shaped, $\eps=1/\sqrt{2}$ (dashed) where it becomes a lemniscate, and $\eps = 0.8$ where it consists of two disconnected corves.}
  \label{fig:cassinoid}
\end{figure}

Unfortunately, this orbit does not fulfill our desire for an equant. The problem is that the reflection principle behind~\eqref{eq:rr} only holds for ellipses, and a curve cannot be both an ellipse and a cassinoid unless it is a circle. Moreover, the cassinoid undergoes some strange transitions in its shape as shown in Figure~\ref{fig:cassinoid}. 

As for the ellipse, define the eccentricity $\eps$ as the distance between the two foci divided by that between perihelion and aphelion. If we again set $a=1$ so the latter distance is $2$, the equation for a cassinoid centered at the origin is
\[
((x+\eps)^2 + y^2)
((x-\eps)^2 + y^2) 
= (1-\eps^2)^2 \, .
\]
For small $\eps$ the cassinoid and the ellipse are fairly similar. For instance, its height at $x=0$ (the semiminor axis) is $\sqrt{1-2\eps^2}$ as opposed to $\sqrt{1-\eps^2}$ for the ellipse. But at $\eps=1/\sqrt{3}$ the cassinoid ceases to be convex and becomes peanut-shaped. At $\eps = 1/\sqrt{2}$ it becomes a lemniscate (i.e., a figure $\infty$), and for larger $\eps$ it consists of two separate lobes. Clearly if we want to satisfy the second law while having an equant, we must look elsewhere.

\section{A Hybrid Universe}

\subsection{Conserving $\omega$ and $J$}

We now enter an alternate universe that only a historian of physics could love. Is there an orbit that obeys Kepler's second law, with a central force that conserves angular momentum around the sun, but where some point is an exact equant as opposed to an approximate one? Can Kepler stand at the sun and observe constant angular momentum $J$, while Ptolemy stands at the equant and observes constant angular speed $\omega$, both astronomers confirmed in their beliefs?

There are several degenerate solutions where $\omega=0$ or $J=0$. A planet can move on a straight line at uniform speed through the equant, or on a straight line through the sun at a time-varying speed that keeps $\omega$ constant. And of course if the sun and the equant are identical, the planet can follow a uniform circular orbit. If these were the only solutions then our hybrid universe would be a disappointing place to visit.

\begin{figure}
  \centering
  \includegraphics[width=2.5in]{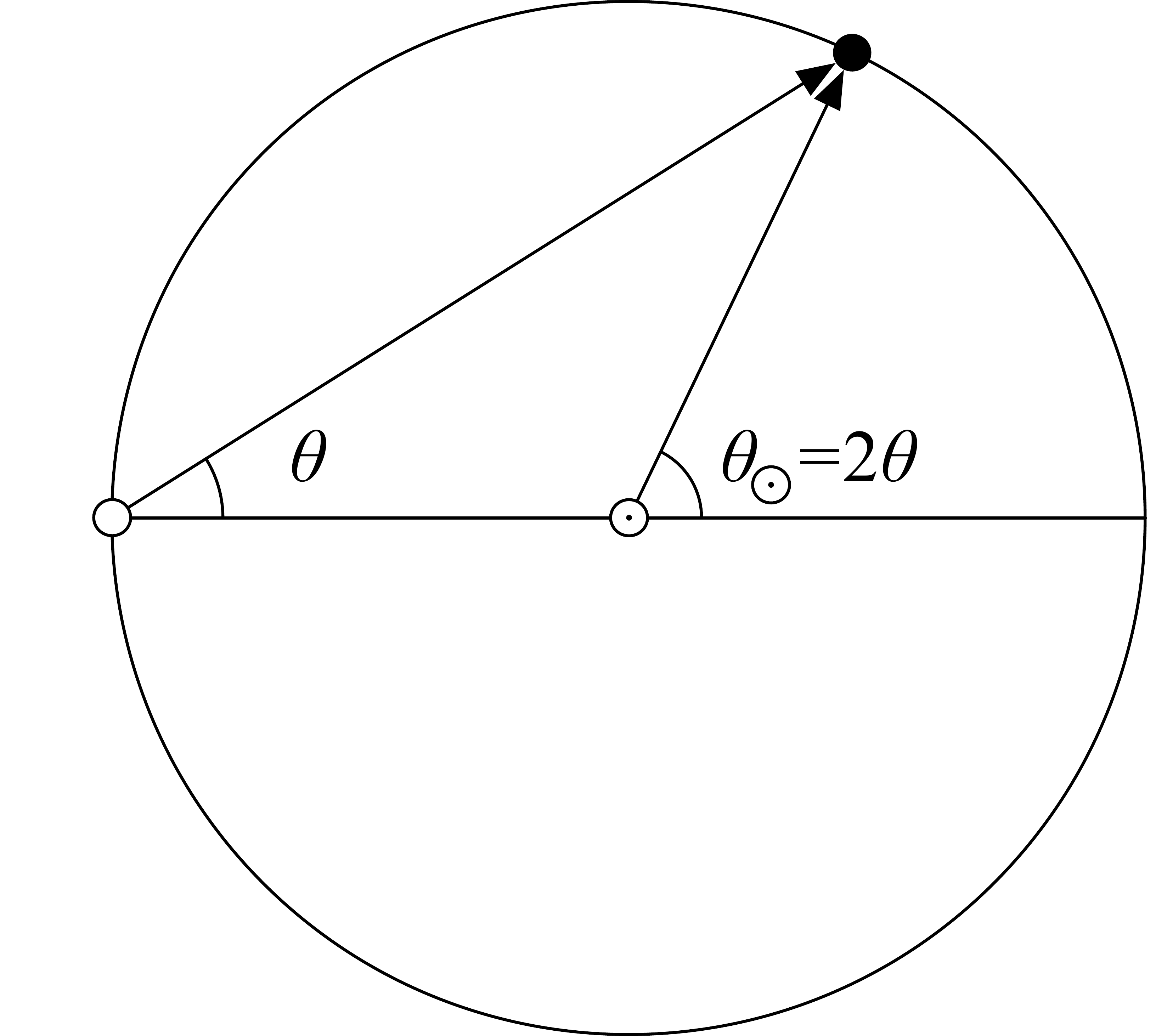}
  \caption{A uniform circular orbit around the sun has constant angular speed around the equant if it passes through it.}
  \label{fig:circular}
\end{figure}

A more interesting solution is a uniform circular orbit around the sun that passes \emph{through} the equant as shown in Figure~\ref{fig:circular}. The angle $\theta$ and angular speed $\omega$ is undefined when $r=0$ and $\theta = \pm \pi/2$. But at all other points $\omega$ is constant since, by elementary geometry, the angle $\theta_\odot$ around the sun is twice the angle around the equant. Thus the planet completes two orbits around the sun for every ``orbit'' around the equant. 


While this orbit is not what we are looking for, it suggests that nontrivial solutions exist for certain values of $J$. To look for these, let's place the equant at the origin $(0,0)$, the sun at $\vec{\ell}=(\ell,0)$, and the planet at $(x,y)$. We then have two differential equations, one for Ptolemy and one for Kepler:
\begin{align} 
\vec{r} \times \vec{v} &= x y' - y x' = \omega r^2 \label{eq:system1} \\ 
(\vec{r}-\vec{\ell}) \times \vec{v} &= (x-\ell) y' - y x' = J \label{eq:system2} 
\, .
\end{align}
Solving these gives 
\begin{align} 
y' &= \frac{\omega r^2 - J}{\ell} \label{eq:y0} \\
x' &= \frac{\omega r^2 (x-\ell) - Jx}{\ell y} \label{eq:x0} \, ,
\end{align}
where $r^2=x^2+y^2$. 
The equation for $y'$ is tantalizingly simple. However, $x'$ has a singularity when $y=0$. Setting its numerator to zero we find three places where $\dx/\dt$ can avoid diverging on the $x$-axis, namely $x=0$ and the roots of the quadratic equation $\omega x (x-\ell) = J$:
\begin{equation}
\label{eq:x-axis}
x_\pm = \frac{\ell}{2} \pm \sqrt{\frac{\ell^2}{4} + \frac{J}{\omega}} \, . 
\end{equation}
These are also the values of $x$ where $y'=\omega x$ and $J=y'(x-\ell)$. 
Thus, unless the planet crosses the $x$-axis at the equant, i.e., at the origin, its perihelion and aphelion are at $x_+$ and $x_-$ respectively. Applying l'H\^{o}pital's rule to~\eqref{eq:x-axis} with $y=0$ and $x \ne 0$ we have 
\[
\ell x' = (2x-\ell) x' \, ,
\]
so except in the degenerate cases $x=0$ or $x=\ell$ we have $x'=0$. Similarly, differentiating~\eqref{eq:system2} gives $y''=0$ when $y=0$. Thus in the nondegenerate case we have $x'=y''=0$ at the $x$-axis as we would expect from symmetry. 

The planet's acceleration $x''$ at perihelion and aphelion is also fixed.  Differentiating~\eqref{eq:system1} and~\eqref{eq:system2} twice and setting $y=x'=y''=0$ gives
\begin{align*}
x y''' - y' x''  
&= 2 \omega (x x'' + y'^2) \\
 (x-\ell) y''' - y' x ''  
 &= 0 \, .
\end{align*}
Eliminating $y'''$ and setting $x=x_\pm$ then gives
\begin{equation}
\label{eq:x-double-prime}
 x'' 
= \frac{2 y'^2 (x-\ell)}{3x\ell-2x^2} 
= \frac{2 \omega J}{3\ell-2x}
\, . 
\end{equation}


This makes the dynamics on the $x$-axis quite singular, although to be fair the same is true for Kepler's laws if we fix the foci and the angular momentum. As we will see next, this hybrid universe has a close relationship to a well-known family of polynomials.

\subsection{Periodic solutions from Legendre polynomials}
\label{sec:periodic}

In Section~\ref{sec:kepler}, we studied Kepler orbits geometrically by rescaling time and space,  setting the time of revolution of a given orbit to $T=2\pi$ so that its mean angular speed is $\overline{\omega}=1$. Similarly, here we rescale time and space so that $\omega=1$ and $\ell=1$. Fixing the distance $\ell$ between the sun and the equant lets us compare orbits of different sizes and different angular momenta $J$. However, as before, we caution the reader that this is not a choice of physical units; it simply lets us avoid carrying $\ell$ and $\omega$ throughout our calculations, so we can focus on the geometric shape of our orbits.

Since $d\theta/\dt = \omega = 1$, we have
\[
\theta = t \bmod 2\pi \, . 
\]
As a caveat, while the orbits in this section will be periodic with period $T=2\pi$, others below will not be. In a periodic orbit, $r(t)$ only contains trigonometric functions of $t$ and is thus a function of $\theta$. But in other orbits, $t$ will  appear as a secular variable in $r(t)$, so successive loops are not the same. Moreover, we will encounter some highly nonphysical orbits where $r(t)$ diverges at finite $t$, which means they fly to infinity in finite time. Nevertheless, these unphysical solutions will help us build the closed curves we are looking for. 

Setting $\omega=\ell=1$ in~\eqref{eq:y0} and writing $r=y\, / \sin \theta=y\, / \sin t$ gives
\begin{equation}
\label{eq:y1}
\frac{\dy}{\dt} = \frac{y^2}{\sin^2 t} - J \, . 
\end{equation}
We can get rid of the $\sin^2 t$ term by transforming variables to $z = \cot t$. Since $\dz/\dt = -1/\sin^2 t = -(1+z^2)$, this gives
\begin{equation}
\label{eq:y2}
\frac{\dy}{\dz} = - y^2 + \frac{J}{1+z^2} \, . 
\end{equation}
Next, we use a classic substitution, 
\[ 
y = (\ln u)' = u'/u \, . 
\]
Since $y' = u''/u - (u')^2/u^2 = u''/u - y^2$, we have
\begin{equation}
\label{eq:u0}
(1+z^2) u'' = J u \, ,
\end{equation}
a linear second-order differential equation.

This equation can be solved using hypergeometric functions, but these solutions are not particularly illuminating. We would prefer polynomial solutions $u(z)$ since they yield periodic solutions $u(\cot t)$ to our original problem. But as for many second-order differential equations, polynomial solutions only exist for certain $J$. To put this more deeply, if we think of $(1+z^2) (\mathrm{d}/\dz)^2$ as a linear operator on the vector space of power series, then~\eqref{eq:u0} is an eigenvalue equation. This operator preserves the subspace consisting of polynomials, and on that subspace it only has certain eigenvalues $J$.

Because~\eqref{eq:u0} has no first-order term, it is particularly easy to solve with power series. If we set
\begin{equation}
\label{eq:power-series}
u(z) = \sum_{i=0}^\infty a_i z^i \, , 
\end{equation}
then we have the recurrence
\begin{equation}
\label{eq:recurrence}
a_{i+2} = \frac{J-i(i-1)}{(i+1)(i+2)} \,a_i \, .
\end{equation}
This suggests a polynomial which is either even or odd, with nonzero $a_i$ only for even $i$ or for odd $i$. However, we can only obtain a polynomial of finite degree $d$ if $J=d(d-1)$ so that $a_i = 0$ for all $i > d$. Henceforth we will write $J=n(n+1)$ and $d=n+1$, and denote the resulting polynomial $w_n$. The first few of these (up to a multiplicative constant determined by $a_0$ or $a_1$) are
\[
\renewcommand\arraystretch{1.25}
\begin{array}{c|c|l}
n & J & w_n \\ \hline
1 & 2 & \frac{1}{2} + \frac{1}{2} z^2 \\
2 & 6 & \frac{1}{2} z + \frac{1}{2} z^3 \\ 
3 & 12 & \frac{1}{8} + \frac{3}{4} z^2 + \frac{5}{8} z^4 \\
4 & 20 & \frac{3}{8} z + \frac{5}{4} z^3 + \frac{7}{8} z^5 \\
5 & 30 & \frac{1}{16} + \frac{15}{16}  z^2 + \frac{35}{16} z^4 + \frac{21}{16} z^6 
\end{array}
\]
Note that all these coefficients are positive, since we start with a positive constant or linear term and the ratio in the recurrence~\eqref{eq:recurrence} is positive until $i(i-1)=J$ and the series ends.

In fact these $w_n$ are closely related to a classical family of orthogonal polynomials. If we take the derivative of~\eqref{eq:u0} and set $P_n = w'_n$, we get 
\[
(1+z^2) P'' + 2z P' - n(n+1) P = 0 \, .
\]
If we change variables to $x=\i z$, this becomes the Legendre equation of electrostatics and spherical harmonics, 
\begin{equation}
\label{eq:legendre-eq}
(1-x^2) P'' - 2x P' + n(n+1) P = 0 \, .
\end{equation}
Thus we can take $w_n$ such that
\begin{equation}
\label{eq:w-legendre}
w'_n(z) = (-\i)^n P_n(\i z) 
\end{equation}
where $P_n$ is the Legendre polynomial of degree $n$ and the leading factor $(-\i)^n$ is chosen so that $w'_n$ is real. 
A classic fact is that the antiderivative of $P_n$ is a linear combination of $P_{n+1}$ and $P_{n-1}$. This gives
\begin{equation}
\label{eq:antid}
w_n(z) 
= (-\i)^{n+1} 
\int_{-1}^{\i z} P_n(x) \,\dx 
= (-\i)^{n+1} 
\,\frac{P_{n+1}(\i z) - P_{n-1}(\i z)}{2n+1} \, ,
\end{equation}
yielding the solution
\begin{equation}
\label{eq:y-legendre}
y(z) = \frac{w'(z)}{w(z)} = (2n+1) \,\frac{\i P_n(\i z)}{P_{n+1}(\i z) - P_{n-1}(\i z)} \, .
\end{equation}

We can obtain another expression using Rodrigues' formula for the Legendre polynomials~\cite[p.559]{arfken2013mathematical}, 
\[
P_n(x) = \frac{1}{2^n n!} \left( \ddx \right)^{\!n} \!(x^2-1)^n \, .
\]
Along with~\eqref{eq:w-legendre} this gives
\begin{equation}
\label{eq:w-rodrigues}
w_n(z) = \frac{1}{2^n n!} \left( \ddz \right)^{\!n-1} \!(1+z^2)^n \, ,
\end{equation}
which gives the elegant
\begin{equation}
\label{eq:y-rodrigues}
y(z) = \frac{w'(z)}{w(z)} = \frac
{\left( \ddz \right)^{\!n} \!(1+z^2)^n}
{\left( \ddz \right)^{\!n-1} \!(1+z^2)^n} 
= \ddz \log \left( \ddz \right)^{\!n-1} \!(1+z^2)^n \, .
\end{equation}

Whether we use~\eqref{eq:y-legendre} or~\eqref{eq:y-rodrigues}, our  solutions are rational functions of $z=\cot t=\cot \theta$, and are thus periodic.  For $n=1$, i.e., $J=2$, we have
\[
y(z) 
= \frac{w'_2}{w_2} 
= \frac{2z}{1+z^2} \, .
\]
This gives
\begin{equation}
y(\cot t) 
= \frac{2 \cot t}{1+\cot^2 t}
= 2 \cos t \sin t
= \sin 2t \, ,
\label{eq:circular}
\end{equation}
and so
\[
r(t) = \frac{y(\cot t)}{\sin t} 
= 2 \cos t \, ,
\]
which is the circular solution of Figure~\ref{fig:circular}.

More generally, we obtain a smooth periodic solution for any odd $n$. Consider the denominator of $y(z)$ in~\eqref{eq:y-legendre}. The antiderivative~\eqref{eq:antid} of $P_n(x)$ has $n+1$ distinct real roots, namely $x=\pm 1$ and $n-1$ additional roots in the interval $(-1,1)$. If $n$ is odd this antiderivative is even, so these roots are all nonzero. By the Fundamental Theorem of Algebra, these exhaust the roots of $P_{n+1}(x)-P_{n-1}(x)$ in the complex plane. Thus the denominator for imaginary $x=\i z$ is always nonzero, and $y(z)$ has no poles. Furthermore, if $n$ is odd the only imaginary root of the numerator $P_n(\i z)$ is $z=0$, and $y(z)$ is zero only when $z=0$. Since $z=\cot t=0$ when $t=\pm \pi/2$, the orbit crosses the equant at the origin as in the circular case. 

For even $n$, the denominator of $y(z)$ in~\eqref{eq:y-legendre} is odd and has a root at $z=0$, while the numerator is even and nonzero at $z=0$. Thus $y(z)$ has a pole at $z=0$, causing the planet to zoom out to infinity along a vertical line and return from the opposite direction. This vertical line crosses the $x$-axis at $x = y \cot t = yz$. As $z$ tends to zero, $w_n$ is dominated by its first term which is linear in $z$, so $y$ tends to $(\ln z)' = 1/z$ and $x=yz$ tends to $1$. Thus the vertical asymptote of the orbit passes through the sun at $(1,0)$. 

Since this pole occurs at $t = \pm \pi/2$, these even-$n$ ``orbits'' fly off to infinity in finite time. This is obviously unphysical---even more so than those of~\cite{xia} who showed that, in Newtonian gravity, velocities and accelerations can diverge in finite time. Nevertheless, as mathematical solutions they will be an ingredient of the closed orbits we derive below.

For both odd and even $n$, we can ask where these orbits cross the positive $x$-axis where $t=0$. As $t \to 0$ we have $z = \cot t \to \infty$. Since $1/\sin^2 t = 1 + \cot^2 t = 1+z^2$, the radius at $t=0$ is 
\begin{equation}
\label{eq:yz}
\lim_{t \to 0} r(t) 
= \lim_{t \to 0} \frac{y}{\sin t} 
= \lim_{z \to \infty} y \sqrt{1+z^2} 
= \lim_{z \to \infty} y|z| 
\, . 
\end{equation}
When $z$ is large, $w_n$ is dominated by its leading term which is $O(z^{n+1})$, so $y = (\ln w_n)'$ tends to $(n+1)/z$, and $\lim_{t \to 0} r(t) = n+1$. Thus the orbit crosses the $x$-axis at $x=n+1$, and in fact if we plug $\omega=\ell=1$ and $J=n(n+1)$ in to~\eqref{eq:x-axis} we get
\[
x_+ =  \frac{1}{2} + \sqrt{\frac{1}{4} + J} = n+1 \, .
\]

\begin{figure}
  \centering
  \includegraphics[width=2in]{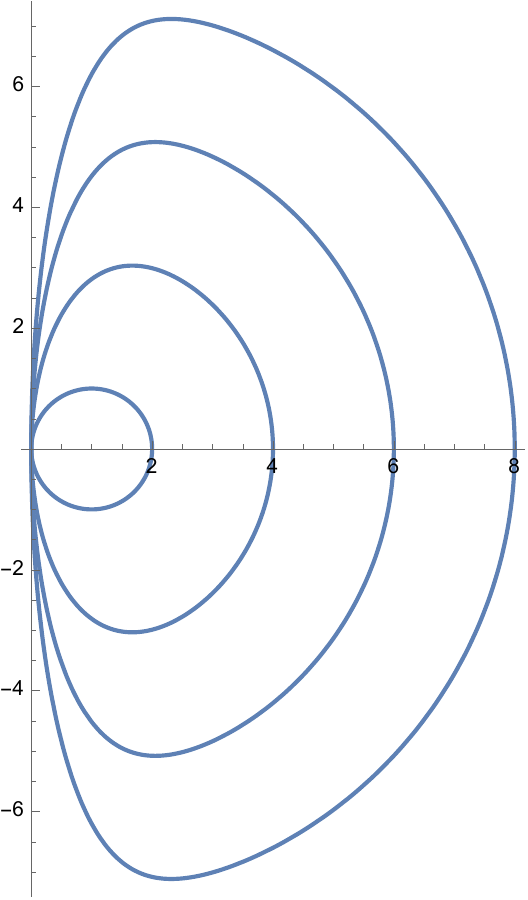}
  \hspace{0.5in}
  \includegraphics[width=2in]{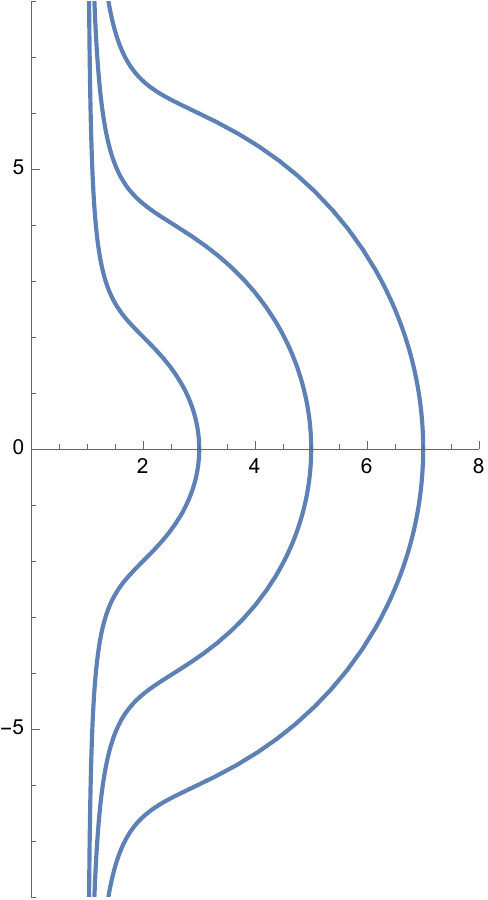}
  \caption{If $J=n(n+1)$ for integer $n$, there is a periodic orbit $r(t)$ that can be written in terms of Legendre polynomials. If $n$ is odd it passes through the equant $(0,0)$. If $n$ is even, $r(t)$ has a pole at $t = \pm \pi/2$ and diverges to infinity in finite time. In both cases the orbit passes through the $x$-axis at $x_+ = n+1$. Here we have $n=1,3,5,7$ (left) and $n=2,4,6$ (right). The sun is at $(1,0)$.}
  \label{fig:orbits}
\end{figure}

We show the orbits for $n=1, 3, 5, 7$ and $n=2, 4, 6$ in Figure~\ref{fig:orbits} on the left and right respectively. As $n$ and $J$ increase, they converge to an elliptical arc interrupted by vertical motion. 

These orbits are lovely, but they are not quite what we wanted. As Ptolemy stands at the equant, he sees the planet moving with uniform angular speed through the sky---but it either disappears to infinity in finite time or obliterates him in an unfortunate collision. Is it really impossible to find an oval orbit that contains both the sun and the equant, just as an ellipse contains its foci?

\subsection{Another family of solutions with the opposite parity}
\label{sec:another}

We have not yet exhausted the solutions of our differential equation~\eqref{eq:u0}. Since it is second order, it has two linearly independent solutions for each $J$. As we have seen, when $J=n(n+1)$ for integer $n$, one solution is the polynomial $w_n$ with even or odd powers of $z$ if $n$ is odd or even respectively. The other solution can also be found in power series form~\eqref{eq:power-series} by applying the recurrence~\eqref{eq:recurrence} to the terms of opposite parity. The resulting series is infinite, yielding a transcendental function which is odd if $w_n$ is even and vice versa.

Let's call these functions $v_n$. They are simply the antiderivatives of the transcendental solutions $Q_n$ of Legendre's equation~\eqref{eq:legendre-eq}. These are known as the Legendre functions of the second kind, and can be written
\[
Q_n(x) = \left(\frac{1}{2} \ln \frac{1+x}{1-x} 
\right) P_n(x) - R_{n-1}(x)
\]
where $P_n$ is the Legendre polynomial and $R_{n-1}$ is a polynomial of degree $n-1$. On the imaginary line $x=\i z$, this becomes
\[
Q_n(\i z) = \i P_n(\i z) \tan^{-1} z 
- R_{n-1}(\i z) \, ,
\]
since, for real $z$,
\[
\frac{1+\i z}{1-\i z} = \e^{2 \i \tan^{-1} \!z} \, .
\]

As in~\eqref{eq:antid}, the antiderivative of $Q_n$ can be written
\begin{equation}
\label{eq:antid-q}
v_n(z) 
= (-\i)^{n+2} 
\,\frac{Q_{n+1}(\i z) - Q_{n-1}(\i z)}{2n+1} \, .
\end{equation}
Here we have chosen the leading constant so that
\begin{equation}
\label{eq:vn}
v_n(z) = w_n(z) \tan^{-1} z + g_n(z) \, ,
\end{equation}
where $w_n(z)$ is our previous solution and $g_n(z)$ is a polynomial of degree $n$. Since 
\[
\tan^{-1} z = z - \frac{z^3}{3} + \frac{z^5}{5} - \frac{z^7}{7} + \cdots
\]
is odd, it both acts as a ``scaffold'' for the infinite series and flips the parity of $w_n$. Since $g_n$ is also odd or even if $n$ is odd or even respectively, $v_n$ and $w_n$ have opposite parity.

The first few $g_n$ are
\[
\renewcommand\arraystretch{1.25}
\begin{array}{c|c|l}
n & J & g_n \\ \hline
1 & 2 & \frac{1}{2} z \\
2 & 6 & \frac{1}{3} + \frac{1}{2} z^2 \\
3 & 12 & \frac{13}{24} z + \frac{5}{8} z^3 \\
4 & 20 & \frac{2}{15} + \frac{23}{24} z^2 + \frac{7}{8} z^4 \\
5 & 30 & \frac{113}{240} z + \frac{7}{4} z^3 + \frac{21}{16} z^5
\end{array}
\]
Like $w_n$, the coefficients of $g_n$ are all positive. To see this, we plug the expression~\eqref{eq:vn} for $v_n$ into our differential equation~\eqref{eq:u0}, use the fact that $w_n$ also satisfies~\eqref{eq:u0}, and find that $g_n$ satisfies the differential equation
\begin{equation}
\label{eq:alt-g}
2 w_n' - \frac{2}{J} \,zw_n'' + (1+z^2) g''_n
= J g_n \, .
\end{equation}
Writing the power series 
\[
w_n 
= \sum_{i=0}^{n+1} 
a_i z^i
\quad \text{and} \quad
g_n 
= \sum_{i=0}^n b_i z^i 
\]
and matching the coefficients of $z^i$, we obtain
\[
2 \left( 1 -\frac{i}{J} \right) (i+1) \,a_{i+1}
+ i(i-1) \,b_i
+ (i+1)(i+2) \,b_{i+2}
= J b_i \, .
\]
Rearranging, we get
\begin{equation}
\label{g-coeffs}
b_{i+2} = \frac{J-i(i-1)}{(i+1)(i+2)} 
\, b_i -
\frac{2(1-i/J)}{i+2} \,a_{i+1} \, . 
\end{equation}
We have $b_{n+2} = 0$, so when $i = n$  the two terms on the right are equal. We showed above that $a_{n+1}$ is positive, so $b_n$ is positive. Proceeding by induction, we see that $b_{n-2}$, $b_{n-4}$, and so on are positive as well. We know that $g_n$ is either an entirely even or entirely odd polynomial, so this covers all non-zero coefficients.

\begin{figure}
  \centering
  \includegraphics[height=4in]{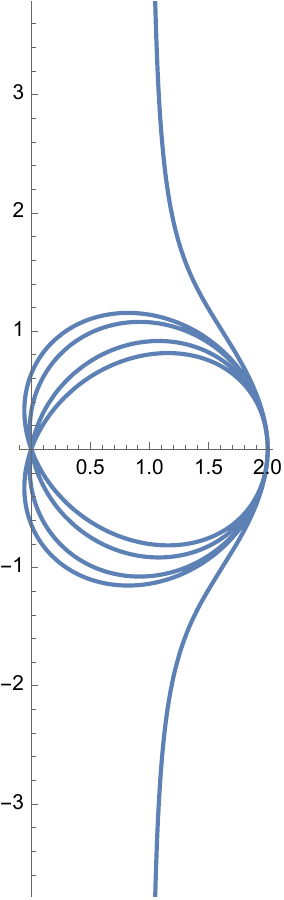}
    \includegraphics[height=4in]{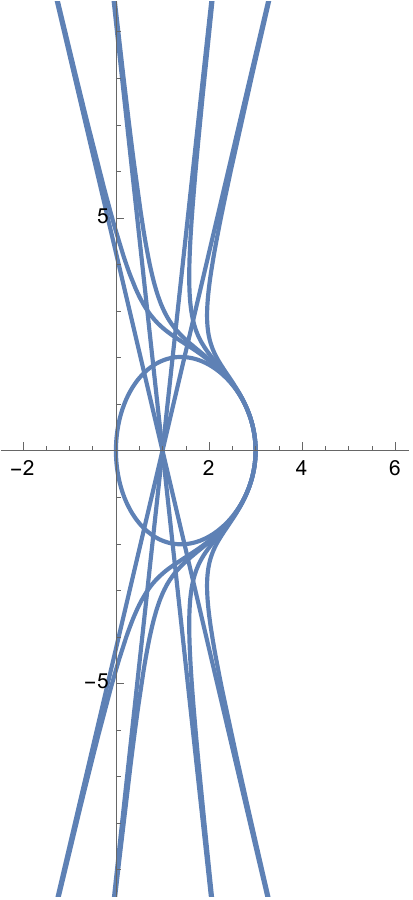}
      \includegraphics[height=4in]{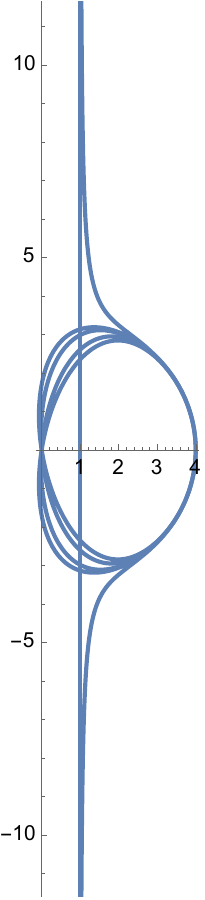}
        \includegraphics[height=4in]{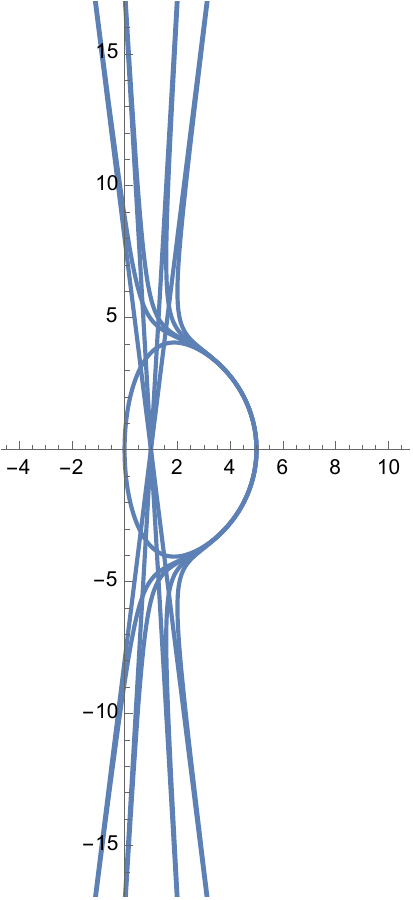}
  \caption{The non-periodic solutions $r(t)$ corresponding to Legendre functions of the second kind from~\eqref{eq:vn} and~\eqref{eq:y-vn} for $n=1,2,3,4$. Each is plotted for $t$ ranging from $-2\pi$ to $3\pi$. Note the convergence to the periodic solutions of Figure~\ref{fig:orbits} as $t \to \pm \infty$.}
  \label{fig:y-vn}
\end{figure}

However, the solutions $y(t)$ corresponding to~\eqref{eq:vn} are not periodic. If we write $y=v'(z)/v(z)$ where $z=\cot t$, the term $\tan^{-1} z = \pi/2 - t$ in~\eqref{eq:vn} makes $t$ appear as a so-called secular variable. That is, $t$ does not just appear inside trigonometric functions, so the orbit varies from one loop to the next. But as $t$ goes to $+\infty$ or $-\infty$, $y(z)$ approaches the periodic orbits described above. To see this, consider (where we suppress the subscript $n$ for readability)
\begin{equation}
\label{eq:y-vn}
y(z) = \frac{v'}{v} 
= \frac{w' \tan^{-1} z + w / (1+z^2) + g'}
{w \tan^{-1} z + g} \, .
\end{equation}
When $t \to \pm \infty$, and thus $\tan^{-1} z \to \pm \infty$, the terms $w' \tan^{-1} z$ and $w \tan^{-1} z$ dominate the numerator and denominator respectively, so $y(z) \to w'/w$.  For instance, for $n=1$ ($J=2$) we have
\[
y(\cot t) = \frac{4 \sin^2 t + (\pi - 2 t) \sin 2t}{\pi + \sin 2t - 2t} \, .
\]
In the limit $t \to \pm \infty$, this approaches the circular orbit $y(\cot t) = \sin 2t$ of Eq.~\eqref{eq:circular}. It also has a pole at $t = \pi/2$, causing it to fly to infinity at finite time like the solutions on the right of Figure~\ref{fig:orbits}. For even $n$ we have a complicated set of poles that converge to the vertical line through the sun as $t \to \pm \infty$. We show the solutions for $n=1,2,3,4$ in Figure~\ref{fig:y-vn}.

\subsection{Orthogonality and the aphelion}

We end this section with an analytic observation. 
The coefficients of $g_n$ are related to those of $w_n$ in the following way:
\begin{equation}
\label{eq:b-a}
b_t = a_{t+1} - \frac{a_{t+3}}{3} + \frac{a_{t+5}}{5} - \frac{a_{t+7}}{7} + \cdots 
\end{equation}
Equivalently, $g_n$ is the ``polynomial part'' of $w_n$ multiplied by the power series
\[
\tan^{-1} \frac{1}{z} = \frac{1}{z} - \frac{1}{3z^3} + \frac{1}{5z^3} - \frac{1}{7z^3} + \cdots \, ,
\]
i.e., the terms of $w_n \tan^{-1} (1/z)$ with nonnegative powers of $z$. Thus $g_n - w_n \tan^{-1} (1/z)$ contains only negative powers of $z$. In fact, the coefficients of the first $n-1$ negative powers of $z$ turn out to be zero, so in the limit $z \to -\infty$ corresponding to $t \to \pi$ we have 
\begin{equation}
\label{eq:surprise}
g_n - w_n \tan^{-1} \frac{1}{z} = O((-z)^{-n}) \, .
\end{equation}
As we will see below, this is essential to our orbit passing through the correct aphelion at $x_ = -n$. 

We can prove~\eqref{eq:surprise} using the orthogonality properties of the Legendre polynomials. First, for an integer $s > 0$, we rewrite~\eqref{eq:b-a} for the coefficient $b_{-s}$ of $z^{-s}$ in $w_n \tan^{-1} (1/z)$. We can use powers of $\i$ to get the pattern of $0$s and $\pm 1$s, giving 
\[
b_{-s} 
= \frac{1}{2\i} \sum_{t=0}^\infty \frac{a_t}{s+t} 
\left( \i^{s+t} - (-\i)^{s+t} \right) \, . 
\]
But since 
\[
\int_{-1}^1 \dz \,(\i z)^{s-1} \,w(\i z) 
= \sum_{t=0}^\infty a_t \int_{-1}^1 \dz \,(\i z)^{s+t-1} 
= \frac{1}{\i} \sum_{t=0}^\infty 
\frac{a_t}{s+t} (\i z)^{s+t} 
\,\Big\vert_{-1}^1 \, ,
\]
we have
\begin{equation}
\label{eq:bs}
b_{-s} 
= \frac{\i^{s-1}}{2} \int_{-1}^1 \dz \,z^{s-1} \,w_n(\i z) \, . 
\end{equation}

Now recall that $w_n(\i z)$ is a constant times the antiderivative of the Legendre polynomial. Rewriting~\eqref{eq:w-legendre} and using the oddness or evenness of $P_n$, we have
\[
w'_n(\i z) = \i (-\i)^n P_n(-z) = \i^{n+1} P_n(z) \, .
\]
This lets us evaluate the intergral in~\eqref{eq:bs} using integration by parts,
\[
\int_{-1}^1 \dz \,z^{s-1} \,w_n(\i z)
= 
\frac{z^s}{s} \,w_n(\i z) \Big\vert_{-1}^1
- \frac{\i^{n+1}}{s} \int_{-1}^1 \dz \,z^s P_n(z) \, .
\]
Since the antiderivative of $P_n$ has roots at $\pm 1$ we have $w_n(\i)=w_n(-\i)=0$, so the boundary term is zero. The second term is the inner product $\langle z^s, P_n \rangle$ where $\langle f, g \rangle = \int_{-1}^1 \dz \,f(z) \,g(z)$. But the Legendre polynomials are orthogonal with respect to this inner product, and $z^s$ is a linear combination of $P_0, P_1, \ldots, P_s$. Thus $\langle z^s, P_n \rangle$ is zero whenever $s < n$ and nonzero when $s=n$, which completes our proof.


\subsection{Closing the loop}

Finally we can present a solution in the form of a closed curve that goes around the sun and the equant, rather than through the equant or off to infinity. We want $r(t)$ to be nonzero for at at least one turn around the origin, and particularly at $t=\pi$.

\begin{figure}
  \centering
  \includegraphics[height=2in]{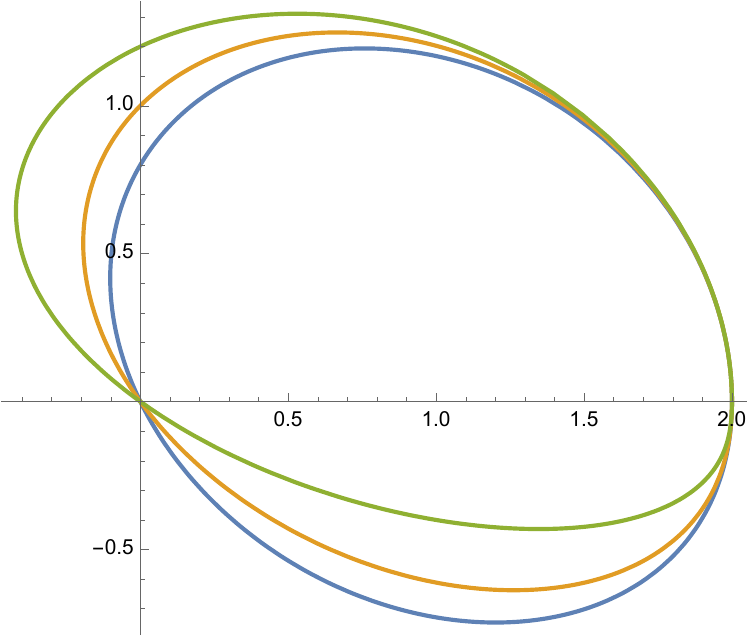}
    \includegraphics[height=3in]{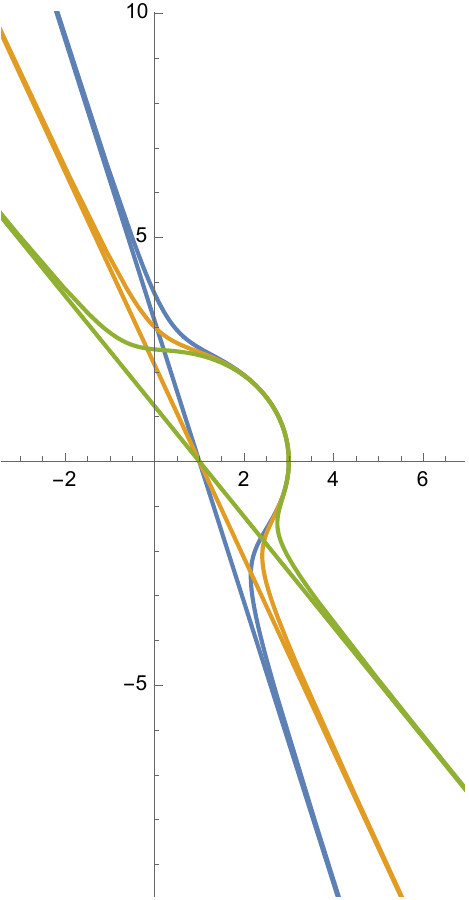}
    \caption{General solutions are given by linear combinations~\eqref{eq:vary} of our two families. As the ratio between their coefficients changes, their zeroes and poles rotate. Here we show $n=1$ and $n=2$ for $\alpha=1$ and $\beta=0.4, 0.5, 0.6$.}
  \label{fig:vary}
\end{figure}

A general solution to~\eqref{eq:u0} is a linear combination of the odd and even solutions, 
\begin{align}
u(z) 
&= \alpha w_n(z) + \beta v_n(z) \nonumber \\ 
&= (\alpha + \beta \tan^{-1} z) w_n + \beta g_n \, . 
\label{eq:vary}
\end{align}
Since $y = (\ln u)' = u'/u$ only the ratio $\alpha/\beta$ matters. As this ratio varies, the zeroes of $u'$ and $u$ move. Whenever $u=0$, $y$ has a pole and diverges; whenever $u'=0$, $y=0$ and it hits the equant. We see this in Figure~\ref{fig:vary} where increasing $\beta$ rotates the solutions from Figure~\ref{fig:orbits}, and their zeroes and poles, counterclockwise.

But if we set $\alpha = \pi/2$ and $\beta=1$, we have
\begin{align*}
u 
&= g_n + (\pi/2 + \tan^{-1} z) w_n \\
&= g_n + (\pi - t) w_n \\
&= g_n - w_n \tan^{-1} (1/z) \, ,
\end{align*}
where we used $z = \cot t = \tan( \pi/2 - t)$. In this case, the zeros of $u$ and $u'$ both tend to $z \to -\infty$, or equivalently $t \to \pi$, allowing a smooth curve for all $z$ and thus for all $t$ between $0$ and $\pi$. 

Recall~\eqref{eq:surprise}, which followed from the orthogonality of the Legendre polynomials. In this limit we have $u=O((-z)^{-n})$, so $y = (\ln u)' = -n \ln (-z) = -n/z = n/|z|$. As in~\eqref{eq:yz}, we have 
\[
\lim_{t \to \pi} r(t) = \lim_{z \to -\infty} y|z| = n \, ,
\]
so these orbits cross the negative $x$-axis at $x = -n$. When $J=n(n+1)$ this is the root $x_-$ given in~\eqref{eq:x-axis}, so perihelion and aphelion occur at $x_+$ and $x_-$ respectively.

\begin{figure}
 \begin{center}
 \includegraphics[width=4in]{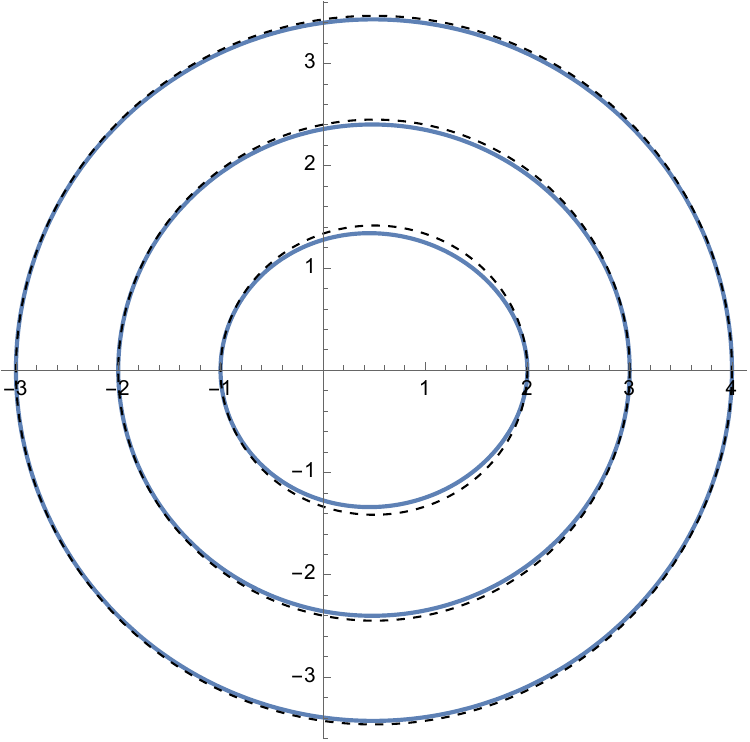}
 \end{center}
 \caption{The closed solutions from Table~\ref{tab:loops} for $n=1,2,3$, compared with Kepler ellipses (dashed) with the same perihelion and aphelion and with the sun and equant as foci.}
 \label{fig:loops}
 \end{figure}

Table~\ref{tab:loops} shows the first few of these solutions for $r(t)$. Figure~\ref{fig:loops} compares these orbits with ellipses with the same perihelion and aphelion, and with foci at the sun and the equant. We note that, in a response to Cassini's proposal described above, in 1702 David Gregory gave an incorrect proof that no convex curve can have both an equant and a point around which the second law holds~\cite{wilson1989predictive}. Our results here provide a family of explicit counterexamples. 

\begin{figure}
 \begin{center}
 \includegraphics[width=3in]{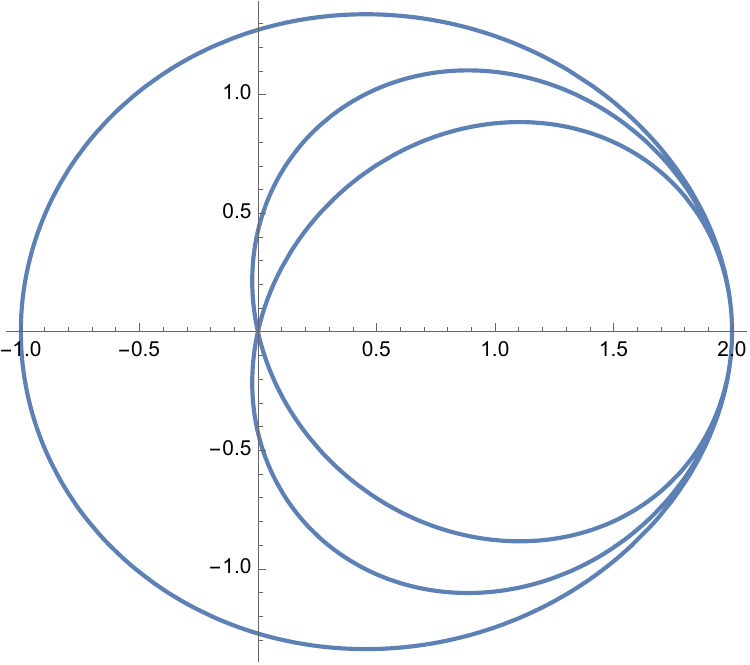}
  \includegraphics[width=3in]{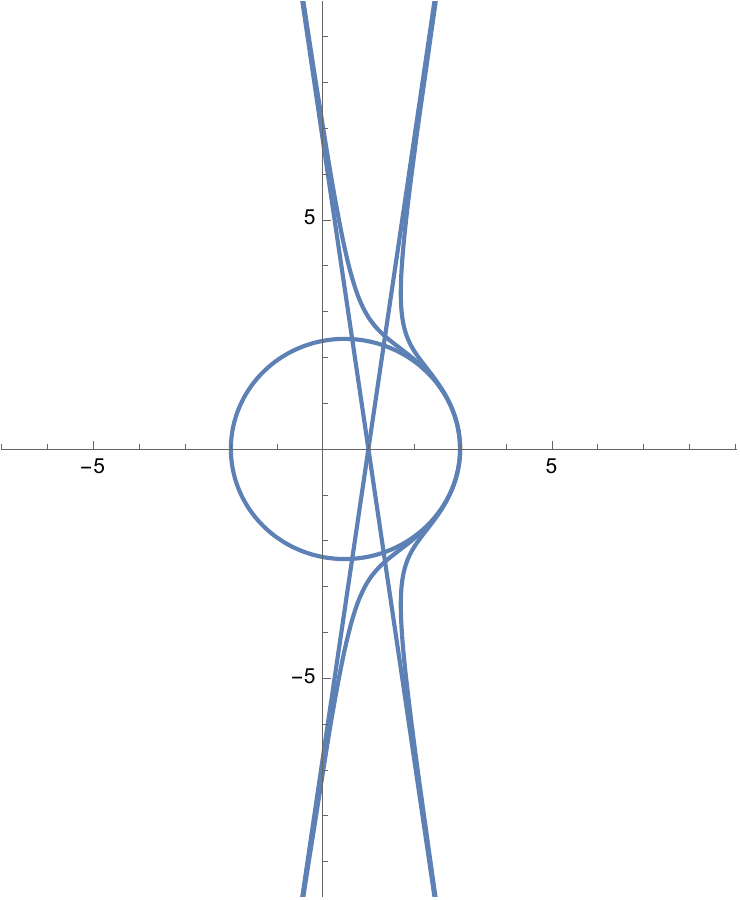}
 \end{center}
 \caption{The closed solutions are not periodic if we extend them analytically outside the interval $0 \le t \le 2 \pi$. Instead, they converge to the solutions from Figure~\ref{fig:orbits}. Here we show the solutions from Table~\ref{tab:loops} for $n=1$ and $n=2$ in the range $-\pi \le t \le 3\pi$.}
 \label{fig:contd}
 \end{figure}

Despite their appearance, these analytic solutions are not periodic, since they still contain $t$ as a secular variable. As shown in Figure~\ref{fig:contd}, if we extend $t$ forwards or backwards beyond the interval $[0,2\pi]$, they converge again to the periodic solutions given in Section~\ref{sec:periodic}. Of course, we can make these solutions periodic by gluing their ends together at perihelion where $r(0)=r(2\pi)$. But their third derivatives are opposite at these two points, so the resulting solutions are only twice-differentiable rather than analytic. Surely Kepler would have asked how the mind of a planet would know when to switch the sign of its third derivative.

\begin{table}
\centering
$
\renewcommand\arraystretch{2.5}
\begin{array}{c|c|c}
n & J & r(t) \\ \hline
1 & 2 & 
\displaystyle{
\frac{2 \sin t + 2 (\pi - t) \cos t)}
{\pi - t +  \sin t \cos t} 
} \\
2 & 6 & 
\displaystyle{
\frac{18 \sin 2t + 12 (\pi -t) (2 + \cos 2t)}
{9 \sin t + \sin 3t + 12 (\pi -t ) \cos t} 
} \\ 
3 & 12 & 
\displaystyle{
\frac{108 \sin t + 44 \sin 3t 
+ 24 (\pi -t)(9 \cos t + \cos 3t)}
   {28 \sin 2t + \sin 4t 
   + 12 (\pi-t) (3 + 2 \cos 2t)} 
 } \\
4 & 20 & 
\displaystyle{
\frac{1600 \sin 2t + 250 \sin 4t 
+  120 (\pi -t ) (18 + 16 \cos 2t + \cos 4t)}
   {300 \sin t + 175 \sin 3t + 3 \sin 5t
   + 120 (\pi -t) (6 \cos t + \cos 3t)}
 } \\
5 & 30 & \displaystyle{
\frac{2000 \sin t + 1625 \sin 3t + 137 \sin 5t
+ 60 (\pi -t ) ( 100  \cos t + 25 \cos 3t + \cos 5t)}
    {425 \sin 2t + 101 \sin 4t + \sin 6t 
    + 60 (\pi -t ) (10 + 10 \cos 2t + \cos 4t)}
 }
\end{array}
$
\caption{Expressions for the closed solutions $r(t)$ shown in Figure~\ref{fig:loops}.}
\label{tab:loops}
\end{table}

\section{Conclusion}

In his efforts to fit the motion of Mars, Kepler tried many possible orbits. These included face-shaped ``metopoids'' and egg-shaped ``composite ovals''~\cite{whiteside-74,davis-09,Carman2025-metopoid}. Like the orbits shown in Figure~\ref{fig:loops}, some of these ovals were curved more sharply at perihelion, and  wider with gentler curvature at aphelion---perhaps because, in his ``vicarious hypothesis,'' he sought an equant as we did here~\cite{donahue93}.  Expressing frustration in Chapter 47 of \emph{Astronomia Nova}, he wrote~\cite{donahue}
\begin{quote}
\ldots we must call in a God from a tragedy–––or rather, a certain reason from a machine---to teach us how to engineer the quadrature of the oval\ldots So now once again, in the oval [ovoid] (or if you prefer, the metopoid), I appeal to geometers and implore their help.
\end{quote}
In Chapter 58, he considered and rejected a ``puffy-cheeked'' orbit, the \emph{via buccosa}~\cite{donahue}:
 \begin{quote}
     It is evident, then, that the path is puffy-cheeked, and therefore not an ellipse. And since the ellipse provides the correct equations, it follows that this puffy-cheeked path [viam buccosam] correctly provides incorrect ones.
\end{quote}
 
Perhaps we should be grateful that Kepler lacked the machinery of differential equations and orthogonal polynomials that we used here, and their ability to satisfy his second law while preserving Ptolemy's equant. As Whiteside wrote~\cite{whiteside-74}, 
\begin{quote}
Looking back upon this omelette of planetary eggs with hindsight and a deeper technical sophistication than Kepler in his day was able to achieve, we must allow that he was several times luckier than he knew---call it the surety of intuition, the unconscious path to truth of the sleepwalker, sheer happenstance, or what you will. Certainly, if he had not persisted stubbornly with his search for the perfect orbit of Mars above and beyond the level of accuracy to which Tycho’s not always exact observations of its solar oppositions gave him the right to attain, he might never have gained his ultimate insight that the ellipse has unique qualities fitting it for the purpose. And if around Easter of 1605 he had been a little less ready spuriously to reject the near-equivalent ‘puffy-cheek’, much future time might sterilely have been spent by himself and later astronomers in pondering its counter-attractions. 
\end{quote}
Now that the danger of distracting Kepler is past, we offer these metopoids to the reader as a fanciful  exploration of what might have been.

\section*{Acknowledgments}

C.M. is supported by the Robert Wood Johnson Foundation. We are grateful to Alex Russell, Christi\'an Carman, Anthony Eagan,  Paola Villa, and Branden Fitelson for helpful conversations. We dedicate this paper to the memory of the third author, who inspired his students with the beauty of physics, astronomy, and cosmology.

\bibliographystyle{alpha}
\bibliography{kepler}

\end{document}